\documentstyle[psfig, epsfig, 12pt]{article} 

\def\mypagenumber{1}
\def\myend{\end{document}}

\normalsize

\newcounter{sxn}

\newcounter{axn}

\date{}

\newdimen\mybaselineskip
\newcommand{\beq}{\begin{equation}}
\newcommand{\eeq}{\end{equation}}
\newcommand{\bea}{\begin{eqnarray}}
\newcommand{\eea}{\end{eqnarray}}
\newcommand{\ba}{\begin{eqnarray}}
\newcommand{\ea}{\end{eqnarray}}
\newcommand{\bpic}{\begin{picture}}
\newcommand{\epic}{\end{picture}}

\def\la{\raise.16ex\hbox{$\langle$} \, }
\def\ra{\, \raise.16ex\hbox{$\rangle$} }

\def\psibar{ \psi \kern-.65em\raise.6em\hbox{$-$} }
\def\mbar{ m \kern-.78em\raise.4em\hbox{$-$}\lower.4em\hbox{} }

\def\n@space{\nulldelimiterspace=0pt \mathsurround=0pt }
\def\huge#1{{\hbox{$\left#1\vbox to 20.5pt{}\right.\n@space$}}}

\def\myskip{\noalign{\kern 8pt}}
\def\myeqspace{\noalign{\kern 10pt}}

\def\boxit#1{$\vcenter{\hrule\hbox{\vrule\kern3pt
    \vbox{\kern3pt\hbox{#1}\kern3pt}\kern3pt\vrule}\hrule}$}
\def\bigbox#1{$\vcenter{\hrule\hbox{\vrule\kern5pt
     \vbox{\kern5pt\hbox{#1}\kern5pt}\kern5pt\vrule}\hrule}$}

\def\ignore#1{{}}

\begin{document}

\bibliographystyle{unsrt}
\footskip 1.0cm

\thispagestyle{empty}
\setcounter{page}{\mypagenumber}

             
\begin{flushright}{\today
\\ OUTP-00-24-P\\
 
}

\end{flushright}

\vspace{1.cm}
\centerline{\Large \bf {Confining strings, Wilson loops and 
}} 

\vspace{0.21cm}

\centerline{\Large \bf {extra dimensions
}}

\vspace{1.5cm}

\centerline{ \bf 
Martin Schvellinger{\footnote{martin@thphys.ox.ac.uk}}}

\vspace{.8cm}

\centerline{\it Theoretical Physics}

\vspace{.2cm}

\centerline{\it Department of Physics}

\vspace{.2cm}

\centerline{\it University of Oxford} 

\vspace{.2cm}

\centerline{\it 1 Keble Road, Oxford, OX1 3NP, UK}


\vspace*{1.cm}


\begin{abstract}
\baselineskip=15pt
We study solutions of the one-loop $\beta$-functions of the critical bosonic string theory
in the framework of the Renormalization Group (RG) approach to string theory, considering
explicitly the effects of the 21 extra dimensions. 
In the RG approach the 26-dimensional manifold is given in terms of 
$M^4 \times R^1 \times {\cal{H}}_{21}$.
In calculating the Wilson loops, as it is well known for
this kind of confining geometry, two phenomena appear: confinement and over-confinement.
There is a critical minimal surface below of which it leads to confinement only.
The role of the extra dimensions is understood in terms of a dimensionless scale $l$ provided
by them. Therefore,
the effective string tension in the area law, the length of the Wilson loops, as well as,
the size of the critical minimal surface depend on this scale. When this 
confining geometry is used to study a field-theory $\beta$-function with an infrared attractive
point (as the Novikov-Shifman-Vainshtein-Zakharov $\beta$-function) the range of the couplings
where the field theory is confining depends on that scale.
We have explicitly calculated the $l$-dependence of that range.
\end{abstract}
\vfill



 
\newpage



\normalsize
\baselineskip=20pt plus 1pt minus 1pt
\parindent=25pt

In a previous paper \cite{HRG} we studied the holographic renormalization group flow
associated to the Yang-Mills theory where the coupling constant diverges in the infrared 
limit\footnote{This has been early studied in references \cite{AL0,ALL}.}
and, $\beta$-functions with an infrared attractive point, as the Novikov-Shifman-Vainshtein-Zakharov (NVSZ)
one \cite{SHIFMAN}. It has been considered
in the framework of the Renormalization Group approach to the string theory
developed by \'Alvarez and G\'omez \cite{AL0,ALL}. In \cite{HRG} it has been calculated the Wilson loops and analyzed
the features of the metric which is a solution of the vanishing string theory sigma model $\beta$-function equations, at leading order in $\alpha'$.
The metric studied shows confinement and also over-confinement, which depend on the size of the minimal
surfaces used to calculate the Wilson loops. In fact, there is a critical minimal surface between those regimes. In the case of 
the NSVZ $\beta$-function, as it was shown in \cite{KOGAN}, there exist
two branches: one behaves asymptotically free in the UV limit and the other one shows a strongly coupled regime in that limit.
There is also an attractive infrared point which means that the RG flow of the theory goes from the UV limit to
the IR limit. In the branch of the strong coupling an interesting result emerges \cite{HRG}, around the 
attractive infrared point, there is a range of couplings where the theory only shows confinement.
From those calculations an open question arises.  That is about the meaning of the extra
$21$-dimensional manifold which is included in order to solve the vanishing one-loop $\beta$-function equations 
of the string theory sigma model in the critical dimension\footnote{In the case of superstrings similar analysis can be done
by means an extra 5-dimensional manifold.}. 
Notice that this solution also satisfies 
the equations of motion of $5$-dimensional gravity coupled to the dilaton field.
In this sense, there is a connection between the RG approach \cite{AL0,ALL} and the 
confining strings \cite{POLYAKOV}\footnote{See also \cite{MAVROMATOS}
where a Liouville-string approach to confinement in four dimensional gauge theories is presented.},
which has been pointed out by \'Alvarez and G\'omez \cite{AL0,ALL}. 

In this paper we study the above mentioned question. In order to
understand the role played by those extra coordinates we will calculate the Wilson loops
using a metric which is a solution of the closed bosonic string theory sigma model $\beta$-function equations, by
considering that the $21$-extra dimensions correspond to a hyper-plane ${\cal{H}}$.
We consider two situations, one is when the Wilson loop is drawn in such a way that the {\it quark} and the
{\it anti-quark} are separated both in the Minkowski spacetime by $L$, and also 
in the hyper-plane ${\cal{H}}$ by a distance $|\Delta {\vec{y}}|$. This point 
will be clarified latter. The other situation, when the {\it quark-antiquark} pair lies at the same point
of the 21-dimensional hyper-plane ${\cal{H}}$ has already discussed in references \cite{AL0,ALL} (and in our context also
in \cite{HRG}). We will use it as a check of consistency in the limit when 
$|\Delta {\vec{y}}| \rightarrow 0$, in the previous case.
The expected result is that the critical size of the minimal surfaces depends on
a scale $l$ which is a function of the separation $|\Delta {\vec{y}}|$ on the $21$-dimensional hyper-plane
\footnote{For the effect of extra dimensions in Wilson loops in the large $N$ limit of field
theories on Calabi-Yau conical singularity see \cite{HERNANDEZ} and references therein.}.
This has an interesting consequence when it is applied to study the NSVZ $\beta$-function 
in the infrared limit (on the branch of strong coupling).
The range of couplings in which the theory is only confining
reduces as $|\Delta {\vec{y}}|$ increases, but as we shall see this reduction is bounded. 
It implies that the effect due
to those extra coordinates leads to a reduction of the range of pure confinement of the theory.  

The first concrete realization of a connection between gravity and field theories
has been made by Maldacena \cite{MALDA}, showing the duality between type IIB supergravity in $AdS_5 \times S^5$
and the ${\cal{N}}=4$ super Yang-Mills theory at the boundary.
This AdS/CFT duality was developed in \cite{MALDA,GUBSER,WITTEN,MALDAREPORT}.
In order to extend the duality to non-conformal field theories it has been proposed several
possibilities \cite{TS}. On the other hand, Polyakov
proposed a string representation of Yang-Mills theories based on the zig-zag invariance of pure Yang-Mills
Wilson loops \cite{POLYAKOV,MIGDAL,POLYAKOV1,KOGAN1,ORTIN}. 
Here we are going to deal with the RG approach to the strings where the idea 
is to model the renormalization group equations of gauge theories from the string theory $\beta$-functions. 
It implies to associate to the couplings
of the gauge field theories some background fields of a closed string theory, so that
the geometry dictates the properties of the gauge theory.

Following reference \cite{AL0,ALL} we solve the vanishing one-loop $\beta$-function equations
of the closed bosonic string theory in the critical dimension. Using the Liouville ansatz \cite{AL0,POLYAKOV} those 
equations are trivially satisfied by any dilaton field. In the metric we identify 4 of the 26 coordinates as the
Euclidean ($\Sigma$) (or Minkowski ($M^4$)) spacetime, where the gauge theory lives. In this prescription it is assumed
the following relation between the string coupling $g_s$, the Yang-Mills coupling $g_{YM}$ and the dilaton:
$g_s=e^\Phi=g^2_{YM}$. The second part of this is related with the soft dilaton theorem \cite{ADEMOLLO,ALL1}.

Let us consider the following metric
\beq
 d s^2 = e^{2 \Phi} (\pm dt^2 + dx_i dx_i) + 4 l_c^2 e^{4 \Phi} (d \Phi)^2  + 
 d{{\vec{y}}_{21}}^2 \,\,\, ,
 \label{METRIC}
\eeq
where $x_i$, $i=1, \, 2, \, 3$, and $l_c$ is an arbitrary scale of dimension of length  
related to the running-energy scale of the field theory considered\footnote{Please notice that in that follows
we will use three different scales: the dimensionless $l$ related to the extra dimensions on ${\cal{H}}$,
the above mentioned $l_c$ (which in \cite{AL0,ALL} is related to the inverse of $\Lambda_{QCD}$), and also
$l_s$ related to the string constant $\alpha'=l^2_s$.}, let us call it $\mu$.
Here $\Phi=\Phi(\mu)$ is the dilaton field.
As we said before we consider a $21$-dimensional hyper-plane ${\cal{H}}$. 
The metric of Eq.(\ref{METRIC}) is a solution of the vanishing one-loop $\beta$-functions of the bosonic strings \cite{CALLAN}
at the leading order in $\alpha'$, which for the critical dimension become 
\beq
 R_{\mu \nu} + 2 \, \nabla_\mu \nabla_\nu \Phi = 0  \,\,\, ,
\eeq
and
\beq
 \nabla^2 \Phi - 2 \, (\nabla \Phi)^2 = 0 \,\,\, ,
\eeq 
where $\nabla_\mu \Phi = \partial_\mu \Phi$ and 
$\nabla_\mu \nabla_\nu \Phi = \partial_\mu \partial_\nu \Phi - 
\Gamma^\alpha_{\mu \nu} \partial_\alpha \Phi$. We set 
the antisymmetric three-form and the tachyon to zero. 
Using the ansatz $\rho=e^{2 \Phi}$ the metric of Eq.(\ref{METRIC}) becomes 
\beq
 d s^2 = \rho \,\, (\pm dt^2 + dx_i dx_i) + l_c^2 \, (d \rho)^2 + d{{\vec{y}}_{21}}^2  \,\,\, ,
\eeq          
where $\rho=\infty$ corresponds to the horizon and $\rho$=0 is the singularity. 
For the region near the singularity the present description breaks down.
Essentially we are dealing with equations of motion for the background
fields at first order in $\alpha'$, {\it i.e.} the background field
approximation. However, as one approaches the singularity
the spacetime becomes strongly curved. In that sense we would need to include
additional terms in the $\alpha'$ expansion for the equations of motion.
It means that we would be considering a limit which is not the
decoupling limit in the sense that $\alpha' \rightarrow 0$ and the
scaled fifth coordinate $\frac{r}{\alpha'}$ is kept fixed. 
Usually, from the point of view of the dual gauge theory, what happens is that
one can solve the weakly coupled limit of the gravity or supergravity
equations of motions and then, through the duality prescription 
\cite{MALDA,GUBSER,WITTEN,MALDAREPORT},
it is possible to know properties of the dual gauge theory. However, there is
an important ingredient in this procedure: the mentioned duality implies
that there is a correspondence between the weakly coupled limit of the gravity
or supergravity theory and the strongly coupled limit of the gauge field theory.
The other way does not work, {\it i.e.}, one can not solve the strongly coupled
limit of gravity by solving first the weakly coupled limit of the gauge theory.
Naively one could think that the singularity would correspond to
a free field theory, however, 
since one can not solve either the theory in the strongly curved space limit, 
it is not possible to give its precise meaning from the point of view of the gauge theory.
We will return to this point at the end of the paper.
\begin{center}
\epsfig{file=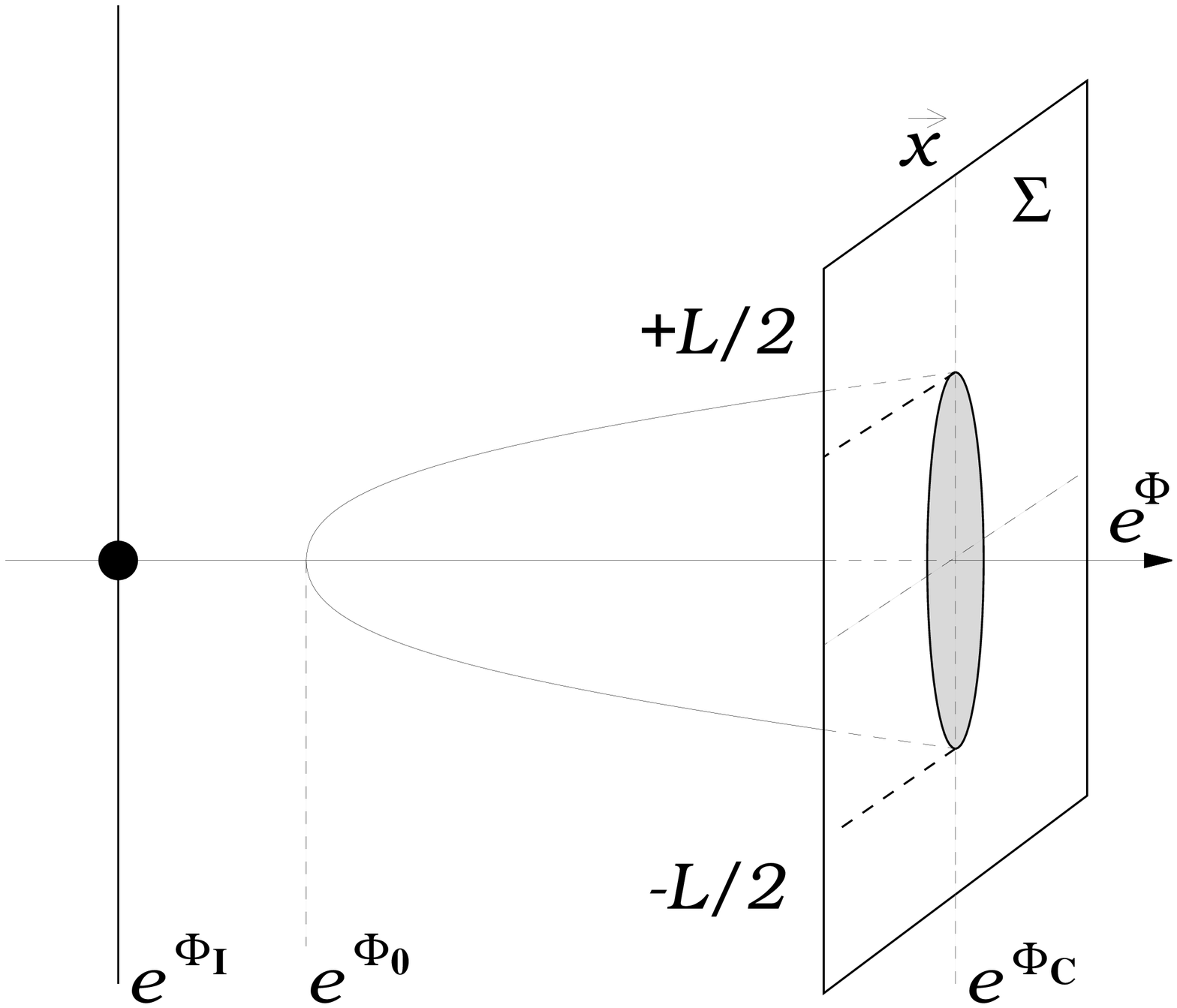, width=8.cm}
\end{center}
\baselineskip=13pt
\centerline{\small{Figure 1: Schematic representation of the prescription of calculating}}
\centerline{\small{the Wilson loops in terms of $e^{\Phi}$-coordinates.}} 
\vspace{1.cm}
\baselineskip=20pt plus 1pt minus 1pt
By computing the expectation value of the Wilson loops it is possible to obtain the energy
of the separation of the {\it quark-antiquark pair}. Let us consider a Wilson loop
${\cal{C}}$, drawn in a $4$-dimensional Euclidean (Minkowski) space-time $\Sigma$, placed at $e^{\Phi_c}$, 
(see figure 1). As it is
usual we consider the limit when the time $T$ is large and use the relation between the
energy of the static configuration $E$ and the action $S$, ($E=S/T$). We follow the Maldacena's proposal 
\cite{MALDAWILSON} for the expectation value of the Wilson loop $<W({\cal{C}})>$.
This expectation value behaves like the exponential of the world-sheet area of a fundamental string
describing the closed curve ${\cal{C}}$ on $\Sigma$ (the curve ${\cal{C}}$ also lies on ${\cal{H}}$
when $|\Delta {\vec{y}}| \neq 0$).

In figure 2 we show a schematic representation of the {\it quark-antiquark}
separation in the hyper-plane ${\cal{H}}$. The figure on the left shows the
case when there is a separation in $x$-coordinates ($L$) which lies on $\Sigma$, but not  
on the hyper-plane ${\cal{H}}$ (here indicated by the vertical $y$-axis). When
the {\it quark} and the {\it antiquark} have different positions at 
the hyper-plane, let us say the points ${\vec{y}}_a$ and ${\vec{y}}_b$, 
the distance $|\Delta {\vec{y}}|$ naturally arises. It is indicated in the figure on the right.

We start from the Nambu-Goto action 
\beq
S_{NG}=\frac{1}{2 \pi l_s^2} \int d\sigma \, d\tau \sqrt{det G_{MN} \partial_\alpha X^M \partial_\beta X^N}
\,\,\, ,
\label{NAMBUGOTO}
\eeq
where $G_{MN}$ is the metric given by Eq.(\ref{METRIC}), and  $X^M$ is a generic coordinate on the $26$-dimensional space-time.
Since we are interested in a static configuration we can take $\tau=t$ and $\sigma=x$. Therefore the metric becomes
\beq
 d s^2 = \pm \, e^{2 \Phi} \, d\tau^2 + 
 (e^{2 \Phi} + 4 \, l_c^2 \, e^{4 \Phi} \, \Phi_\sigma^2 + (\partial_\sigma {\vec{y}})^2) 
 \,\, d\sigma^2  \,\,\, ,
\eeq
where $\Phi_\sigma=\frac{\partial \Phi}{\partial \sigma}$, while 
$\partial_\sigma {\vec{y}}=\frac{\partial {\vec{y}}}{\partial \sigma}$. In addition, let us
remember that $[\tau]=[\sigma]=[l_c]=[l_s]=[y]$ have dimension of length. In the static configuration the action is
\beq
S_{NG}=\frac{T}{2 \pi l_s^2} \int^{L}_0 d\sigma \,  e^{\Phi} \sqrt{e^{2 \Phi} + 4 e^{4 \Phi} l^2_c \Phi_\sigma^2 +
  (\partial_\sigma {\vec{y}})^2} \,\,\, .
\label{NEW}
\eeq

\begin{center}
\epsfig{file=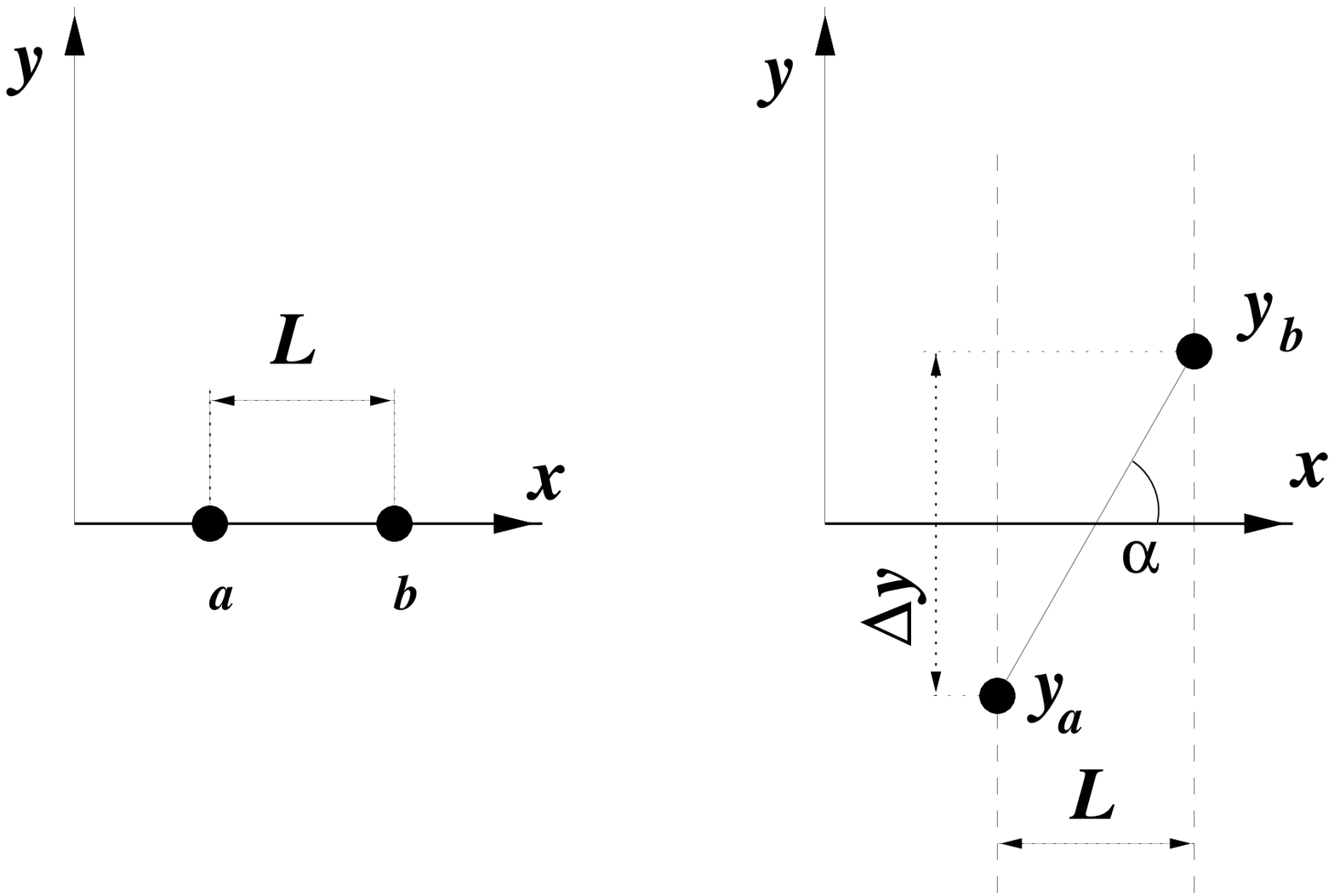, width=10.cm}
\end{center}
\baselineskip=13pt
\centerline{\small{Figure 2.}}
\vspace{0.5cm}
\baselineskip=20pt plus 1pt minus 1pt

We consider configurations as it is shown in figure 1, where $e^{\Phi_I}$ is the limit when $\Phi \rightarrow -\infty $
(represented by a dot at the origin). Since
the scalar curvature $R$ goes like $e^{-4 \Phi}$, it implies that at this point the metric has a naked singularity. 
As we said before in that limit the present description breaks down.
Since $e^{\Phi}=g_s=g_{YM}^2$, it is possible to choice the position of the $4$-dimensional
hyper-plane $\Sigma$ at any point of the $\mu$ coordinate, 
which corresponds to choosing a particular value of the coupling constant, here
denoted by $e^{\Phi_c}$. The string world-sheet can fluctuate orthogonally to the
hyper-plane $\Sigma$, but the minimal surfaces are those corresponding to fluctuations in the direction of
$e^{\Phi_I}$. As we pointed out in \cite{HRG}, in terms of the gauge theory language this means that
semi-classically the RG flow goes from the strongly coupled regime to the weakly coupled one. However, 
if there is an attractive infrared point, it flows
from the ultraviolet limit to the infrared one.
We will consider the vertical axis as the ${\vec{x}}$-direction and study a Wilson-loop of size $L$, as it
is shown in figure 1. 
In the symmetric configuration ${\vec{x}}=0$ corresponds to the minimum $e^{\Phi_0}$, that is when one takes $\Phi_0=\Phi(\mu_0)$. 
For such a minimum $\partial_\sigma {\vec{y}}$ takes the value $\partial_\sigma {\vec{y}}_0$.
Classically the following conditions are straightforwardly obtained by solving the Euler-Lagrange equations for the action of Eq.(\ref{NEW})
\beq
\frac{e^{3 \Phi}}{\sqrt{e^{2 \Phi} + 4 e^{4 \Phi} l^2_c  \Phi_\sigma^2 +
 (\partial_\sigma {\vec{y}})^2}} = e^{2 \Phi_0} \, \sqrt{1-l^2} \,\,\, ,
 \label{WW}
\eeq
and
\beq
\frac{e^{\Phi-\Phi_0}  \, |{\partial_\sigma {\vec{y}}}|}{\sqrt{e^{2 \Phi} + 4 e^{4 \Phi} l^2_c \Phi_\sigma^2 +
(\partial_\sigma {\vec{y}})^2}} =   
 \frac{| \partial_\sigma {\vec{y}}_0|}{\sqrt{e^{2 \Phi_0} + 
  (\partial_\sigma {\vec{y}}_0)^2}}
 =  \, l \,\,\, .
 \label{C2}
\eeq
Those are obtained since the Lagrangian in the above action is independent of ${\vec{x}}$ and ${\vec{y}}$. In this way
Eq.(\ref{WW}) comes from the {\it energy conservation} while Eq.(\ref{C2}) is derived from {\it momentum conservation} in the hyper-plane
${\cal{H}}$.
From these expressions one gets the following relation
\beq
 | \partial_\sigma {\vec{y}}_0|^2 = \frac{l^2 \, e^{2 \, \Phi_0}}{1-l^2} \,\,\, ,
\eeq
where $l$ is related to the distance between the {\it quark-antiquark} pair in the hyper-plane ${\cal{H}}$. In fact $l$
is the $\sin \alpha$, where $\alpha$ is the angle between the {\it quark} and the {\it antiquark} as it is indicated in figure 2.
Notice that from Eq.(\ref{C2}), for  $|\partial_\sigma {\vec{y}}_0| \rightarrow \infty$
we have $l \rightarrow 1$, while $|\partial_\sigma {\vec{y}}_0| << e^{\Phi_0}$ corresponds to $l \rightarrow 0$, 
which means that the distance on the hyper-plane ${\cal{H}}$ has been taken to be very small.
Taking into account the previous conditions of Eqs.(\ref{WW}) and (\ref{C2}), 
we can invert those expressions in order to obtain
\beq
d \sigma = 2 e^{\Phi_0} \, l_c \, \sqrt{1-l^2} \, \frac{dv}{\sqrt{(v^2-1)(v^2-l^2+1)}} \,\,\, ,
 \label{EQUIS}
\eeq
where we use the new variable $v=\frac{e^\Phi}{e^{\Phi_0}}$, while
\beq
d y = 2 e^{2 \Phi_0} \, l \, l_c \, \frac{v^2}{\sqrt{(v^2-1)(v^2-l^2+1)}} \, dv \,\,\, .
\eeq
Notice that for $l=0$ the distance $|\Delta {\vec{y}}|$ trivially vanishes and Eq.(\ref{EQUIS}) 
reduces to the case considered in \cite{HRG}.
Integrating Eq.(\ref{EQUIS}) one obtains the length of the Wilson loop
\bea
\frac{L}{2} & = & 2 \, l_c \, e^{\Phi_0} \, \sqrt{1-l^2} \, \int^{e^{\Phi_c}/e^{\Phi_0}}_1  \, 
 \frac{1}{\sqrt{(v^2-1)(v^2-l^2+1)}} \, dv \nonumber \\
 & = & 2 \, l_c \, e^{\Phi_0} \, \sqrt{1-l^2}
 \, \frac{F \left( \arccos{\left( \frac{e^{\Phi_0}}{e^{\Phi_c}} \right)}, \sqrt{\frac{1-l^2}{2-l^2}} \right)}{\sqrt{2-l^2}} \,\,\, ,
 \label{LLOOP}
\eea
which in the limit of $l \rightarrow 0$ reduces to the case studied in \cite{HRG}.
On the other hand, the distance on ${\cal{H}}$ results
\ba
\frac{\Delta y}{2} &=& 2 e^{2 \Phi_0} \, l_c \, l \, \int^{e^{\Phi_c}/e^{\Phi_0}}_1  \, 
 \frac{v^2}{\sqrt{(v^2-1)(v^2-l^2+1)}} \, dv \nonumber \\
 & = & 2 e^{2 \Phi_0} \, l_c \, l \,  
 \frac{ F \left(\arccos\left( \frac{e^{\Phi_0}}{e^{\Phi_c}} \right), \sqrt{\frac{1-l^2}{2-l^2}} \right)}{\sqrt{2-l^2}} \nonumber \\
 & - & 2 e^{2 \Phi_0} \, l_c  \, l \, \sqrt{2-l^2} \, 
 E \left(\arccos\left( \frac{e^{\Phi_0}}{e^{\Phi_c}} \right), \sqrt{\frac{1-l^2}{2-l^2}} \right) \nonumber \\
 & + & 2  \, l_c \, l \, e^{3 \Phi_0-\Phi_c} \,
 \sqrt{(e^{2(\Phi_c-\Phi_0)}+1-l^2)(e^{2(\Phi_c-\Phi_0)}-1)}
 \label{LY} 
 \,\,\, ,
\ea
which trivially reduces to $0$ as $l \rightarrow 0$, as it is expected from \cite{HRG}.
Here $F$ and $E$ are elliptic integrals of the first and second kind \cite{GRA}, respectively.

\vspace{0.5cm}
\begin{center}
\epsfig{file=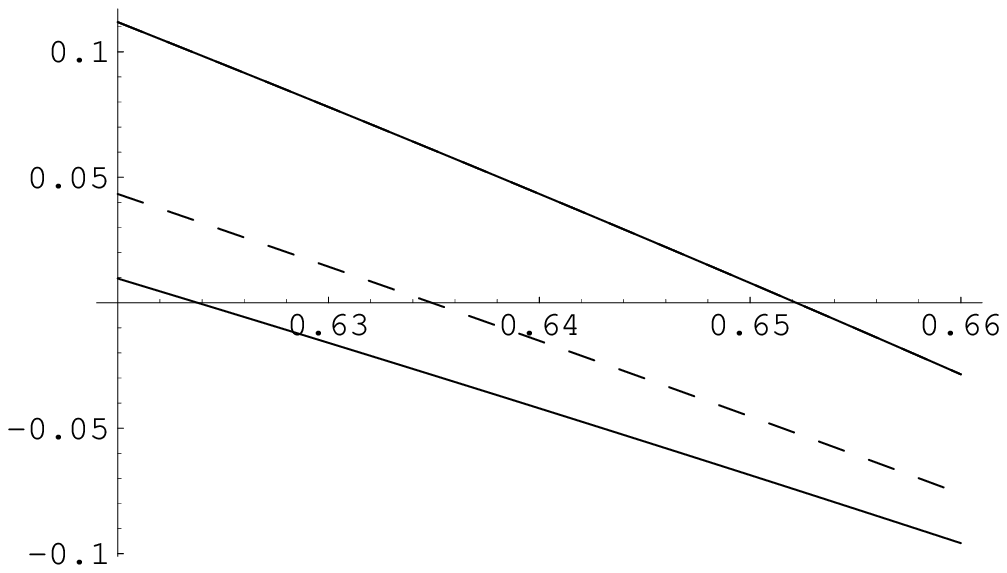, width=9cm}
\end{center}
\vspace{0.5cm}
\baselineskip=13pt
\centerline{\small{Figure 3: Derivative of the scaled Wilson-loop size $L/e^{\Phi_c}$ (in units of $l_c$) in terms of the}} 
\centerline{\small{dimensionless variable $e^{\Phi_0}/e^{\Phi_c}$. The zero of that derivative (maximum of $L/e^{\Phi_c}$) moves to}}
\centerline{\small{the right as $l \rightarrow 1$. The zeros of the solid lines correspond to the extreme values of}}
\centerline{\small{$l=0$ (left, lower solid straight line) and 1 (right, upper solid straight line), respectively.}}

\vspace{0.5cm}

\baselineskip=20pt plus 1pt minus 1pt
From Eqs.(\ref{LLOOP}) and (\ref{LY}) we see that as $l$ approaches to one,
the separation of the pair on $\Sigma$ decreases, while the separation in ${\cal{H}}$ increases.
We will come back on this point latter.
Since $L$ is function of $e^{\Phi_c}$ and $e^{\Phi_0}$, we can plot $L/e^{\Phi_c}$ as a
function of $e^{\Phi_0}/e^{\Phi_c}$. There is a maximum between the two regions (I and II) \cite{HRG,AL0,ALL,REY}. 
The region I, {\it i.e.} below the maximum, corresponds to large world-sheets.

In figure 3 we plot the derivative of $L/e^{\Phi_c}$ as a function 
of $e^{\Phi_0}/e^{\Phi_c}$. We show three curves, the lower one (solid line) corresponds to $l=0$.
The maximum of $L/e^{\Phi_c}$ occurs at $e^{\Phi_0^M}/e^{\Phi_c} \approx 0.62$ and it is $L/e^{\Phi_c}|_{Maximum}=0.42 \, \, l_c$.
It is precisely what happens when one reduces the problem to consider the {\it quark} and the {\it antiquark} at the same point in ${\cal{H}}$
\cite{HRG,AL0,ALL}. On the other hand,
we also plot the cases when $l=0.7$ (showed by the dashed line) and $l=1-10^{-15}$ (this is the upper solid line).
The size of both the critical minimal surface and the Wilson loop $L$ decreases
as $l \rightarrow 1$. For instance, when we take
$l=1-10^{-15}$, we are considering the limit of large separation on the $21$-dimensional hyper-plane
(see for instance figure 4). In this situation  
$e^{\Phi_0^M}/e^{\Phi_c}$ is approximately $0.652$.
Therefore, the effect of the extra dimensional manifold ${\cal{H}}$ is reflected on the size of the minimal surfaces and the corresponding
Wilson loop sizes.  

\vspace{0.5cm}
\begin{center}
\epsfig{file=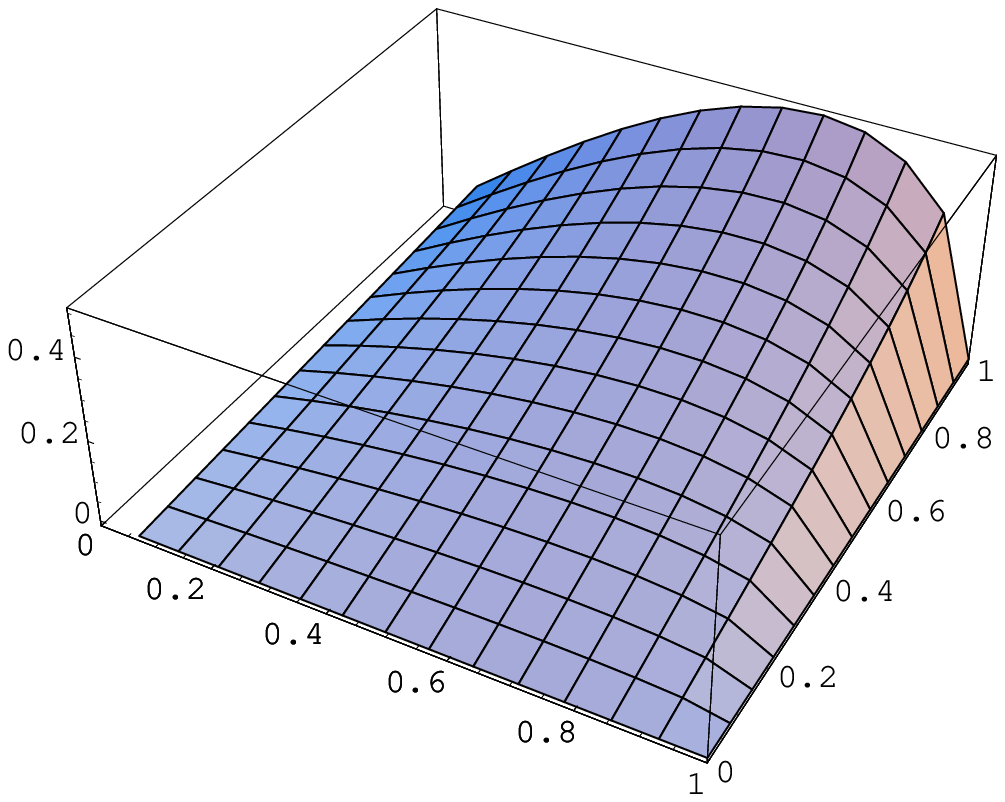, width=9cm}
\end{center}
\vspace{0.5cm}
\baselineskip=13pt
\centerline{\small{Figure 4: Plot of $|\Delta {\vec{y}}|/e^{2\Phi_c}$ (in units of $l_c$) as a function of the dimensionless}} 
\centerline{\small{variable $e^{\Phi_0}/e^{\Phi_c}$ (horizontal axis) and $l$ (transverse horizontal axis).}}

\vspace{0.5cm}

\baselineskip=20pt plus 1pt minus 1pt

By using the RG approach to the strings \cite{AL0,ALL}, if one considers a $\beta$-function with a pole in the infrared
\cite{HRG,KOGAN}, there is a range of couplings $(g_{YM}^\Lambda, \,\, 0.62^{-1/2} \, g_{YM}^\Lambda \, (\equiv 1.27 \, g_{YM}^\Lambda))$ for $l=0$, and
$(g_{YM}^\Lambda, \,\, 0.65^{-1/2} \, g_{YM}^\Lambda \, (\equiv 1.24 \, g_{YM}^\Lambda))$ 
for $l=1$, where only confinement takes place. Here $\mu=\Lambda$ is the infrared attractive point.
 
In figure 4 we plot $|\Delta {\vec{y}}|/e^{2\Phi_c}$ (in units of $l_c$) as a function of 
$e^{\Phi_0}/e^{\Phi_c}$ and $l$. For $l=0$, $|\Delta {\vec{y}}|/e^{2\Phi_c} = 0$. For 
finite values of $l$ the maximum moves up to $e^{\Phi_0}/e^{\Phi_c}=0.652$, which corresponds to the limit for $l=1$.
Let us calculate the energy for the static configuration. 
\bea
E  &=&  2 \, e^{3 \Phi_0} \, \frac{l_c}{\pi \, l^2_s} \, (1+l^2) \, 
 \frac{F\left( \arccos{\left( \frac{e^{\Phi_0}}{e^{\Phi_c}} \right)}, \sqrt{\frac{1-l^2}{2-l^2}} \right)}{3 \sqrt{2-l^2}} 
 \nonumber \\
  & - & 4 \, e^{3 \Phi_0} \, \frac{l_c}{\pi \, l^2_s} \,  l^2 (2-l^2) \,
\frac{E\left( \arccos{\left( \frac{e^{\Phi_0}}{e^{\Phi_c}} \right)}, \sqrt{\frac{1-l^2}{2-l^2}} \right)}{3 \sqrt{2-l^2}} 
\nonumber \\
  & + & \frac{2 \, l_c}{3 \pi \, l^2_s} \, e^{4 \Phi_0-\Phi_c} \,
 (2 l^2 + e^{2(\Phi_c-\Phi_0)}) \, \sqrt{(e^{2(\Phi_c-\Phi_0)}-l^2+1)(e^{2(\Phi_c-\Phi_0)}-1)} \,\, .
 \label{ENERGY1}
\eea
Expanding Eqs.(\ref{LLOOP}) and (\ref{ENERGY1}) in series of 
powers of $e^{\Phi_0}/e^{\Phi_c}$ we get
\beq
\frac{L}{l_c}  = 4 \, e^{\Phi_0} \sqrt{1-l^2} \, \frac{K\left( \sqrt{\frac{1-l^2}{2-l^2}} \right)}{\sqrt{2-l^2}} - 
 4 \, \sqrt{1-l^2} \, e^{2 \Phi_0-\Phi_c} + {\cal{O}}(e^{4 \Phi_0 - 3 \Phi_c})  
  \,\,\, ,
  \label{LL2}
\eeq
and
\bea
  \frac{\pi l^2_s E}{2 l_c} & = & \frac{e^{3 \Phi_c}}{3} + \frac{l^2}{2} \, e^{\Phi_c+2 \Phi_0}  \nonumber \\
  &+& \left( -2 \, l^2 \sqrt{2-l^2} \, E\left( \sqrt{\frac{1-l^2}{2-l^2}} \right) + (1+l^2) \, 
  \frac{K\left( \sqrt{\frac{1-l^2}{2-l^2}} \right)}{\sqrt{2-l^2}} \right)  \,
    \frac{e^{3 \Phi_0}}{3} \nonumber \\
  &+& {\cal{O}}(e^{4\Phi_0-\Phi_c})
  \,\,\, .
\label{ENERGY2}
\eea
Here $K(t)=F(\frac{\pi}{2},t)$ and $E(t)=E(\frac{\pi}{2},t)$ are the complete elliptic integrals of the
first and second kind, respectively.
Replacing $L$ in terms of $e^{\Phi_0}$ in Eq.(\ref{ENERGY2}) we get 
\beq
E = \frac{1}{96 \pi l^2_s l^2_c} \, \left(\sqrt{\frac{2-l^2}{1-l^2}}\right)^3 \, 
  \frac{\left( - 2 \, l^2 \, (2-l^2) \, E\left( \sqrt{\frac{1-l^2}{2-l^2}} \right) +
  (1+l^2) \, K\left( \sqrt{\frac{1-l^2}{2-l^2}} \right) \right)}{ \sqrt{2-l^2} \, K^3\left( \sqrt{\frac{1-l^2}{2-l^2}} \right)} \, L^3
  + {\cal{O}}({e^{3 \Phi_c}}) \,\,\, ,
    \label{gap}
\eeq
which leads to over-confinement.
In the strongly coupled limit ($e^{\Phi_c} \rightarrow \infty$) the above expression becomes
divergent. However, the divergent part is independent of the size of the Wilson loop. 
In the context of AdS/CFT duality this kind of divergence was regularized by a mass renormalization \cite{MALDAWILSON,REYY} or by considering
the Legendre transform of the minimal area \cite{GROSS}.   
In our case, observing that zero-size Wilson loops diverge we will drop the $L$ independent divergent term  
and measure the energy with respect to the zero-size Wilson loops. 
In this limit $L$ becomes
\beq
L_{\infty} = 4 \, l_c \, e^{\Phi_0} \, \sqrt{\frac{1-l^2}{2-l^2}}
 \, K \left(\sqrt{\frac{1-l^2}{2-l^2}} \right) \,\,\, ,
 \label{LINFINITY}
\eeq
which is exactly the same as in \cite{HRG} when one takes $l=0$, that is 
\beq
L^0_{\infty} = \sqrt{\pi} \, l_c \, e^{\Phi_0} \, \frac{\Gamma(\frac{1}{4})}{\Gamma(\frac{3}{4})} \,\,\, .
\eeq
The energy for that configuration is
\beq
E_\infty  =  2 \, e^{3 \Phi_0} \, \frac{l_c}{\pi \, l^2_s} \, (1+l^2) \, 
 \frac{K\left(\sqrt{\frac{1-l^2}{2-l^2}} \right)}{3 \sqrt{2-l^2}} 
   - 4 \, e^{3 \Phi_0} \, \frac{l_c}{\pi \, l^2_s} \,  l^2 (2-l^2) \,
\frac{E\left(\sqrt{\frac{1-l^2}{2-l^2}} \right)}{3 \sqrt{2-l^2}} 
\,\, .
 \label{ENERGYINF}
\eeq
In the limit $l=0$ it reduces to
\beq
E_{\infty} = \frac{2 l_c \, e^{3 \Phi_0}}{3 \sqrt{2} \pi l^2_s} K\left( \frac{1}{\sqrt{2}} \right) = 
 \frac{l_c \, e^{3 \Phi_0}}{6 \sqrt{\pi} l^2_s} \frac{\Gamma(\frac{1}{4})}{\Gamma(\frac{3}{4})} \,\,\, ,
\eeq
which is, as it is expected, proportional to $L^3$.

A quite different situation arises when the region II is analyzed.  In this region
$e^{\Phi_0}$ is close to $e^{\Phi_c}$. In such a case we have to do the integration between
$1$ and $1+\epsilon$, so that we have
\beq
\frac{L_\epsilon}{2}  = 
  2 l_c \, e^{\Phi_0} \, \sqrt{\frac{1-l^2}{2-l^2}} \, \left( \sqrt{2} \, \epsilon^{1/2} -  \frac{5}{6 \, \sqrt{2}} \, \epsilon^{3/2}
   + {\cal{O}}(\epsilon^{\frac{5}{2}}) \right) \,\,\, ,  
\eeq
and
\beq
E_\epsilon  = 
  \frac{2 \sqrt{2}  l_c}{\pi l^2_s} \, e^{3 \Phi_0} \frac{1}{\sqrt{2-l^2}} \, \sqrt{\epsilon}
     + {\cal{O}}(\epsilon^{\frac{3}{2}}) 
 \,\,\,  .
\eeq
At order $\sqrt{\epsilon}$ it gives 
\beq
E_\epsilon = \frac{e^{2 \Phi_0}}{2 \pi l^2_s} \, \frac{1}{\sqrt{1-l^2}} \, L_\epsilon \,\,\, ,
\label{LINEAL} 
\eeq
indicating confinement at region II (area law). The above expression reduces to the case of \cite{HRG} when $l=0$. In the present
context the extra dimensions increase the value of the effective tension by a factor $\frac{1}{\sqrt{1-l^2}}$,
which is expected since it takes into account the separation of the {\it quark-antiquark} pair on ${\cal{H}}$.
The separation $|\Delta {\vec{y}}|/e^{2\Phi_c}$ in the hyper-plane ${\cal{H}}$ also behaves linearly as 
a function of $L$ and it is an increasing function of $l$ as it is shown in figure 4. For every value of $l$ the maxima of 
$|\Delta {\vec{y}}|/e^{2 \Phi_c}$ and $L/e^{\Phi_c}$ take place at the same point, {\it i.e.} $e^{\Phi^M_0}/e^{\Phi_c}$, as it should be expected.

In figure 5 we show the shape of critical minimal surfaces for different values of $l$. The hyper-plane $\Sigma$ is placed at $e^{\Phi_c}$.
The solid line represents the critical minimal surface corresponding to $l=0$. The dotted lines represent critical
minimal surfaces for $0<l<1$. Given some value of $l$
there is a critical minimal surface below of which the theory shows confinement only. Above of that 
there is over-confinement.
In the particular case of the NSVZ $\beta$-function, it has a pole in the infrared, it means that given a value of the coupling,
{\it i.e.} $g^2_{YM}=e^{\Phi_c}$, the size of the critical minimal surface is upper bounded. It implies that the range of couplings
where the theory has only confinement decreases as $l$ increases from $g_{YM}= 1.27 \, g_{YM}^\Lambda$ 
to $g_{YM}= 1.24 \, g_{YM}^\Lambda$.

\vspace{0.5cm}
\begin{center}
\epsfig{file=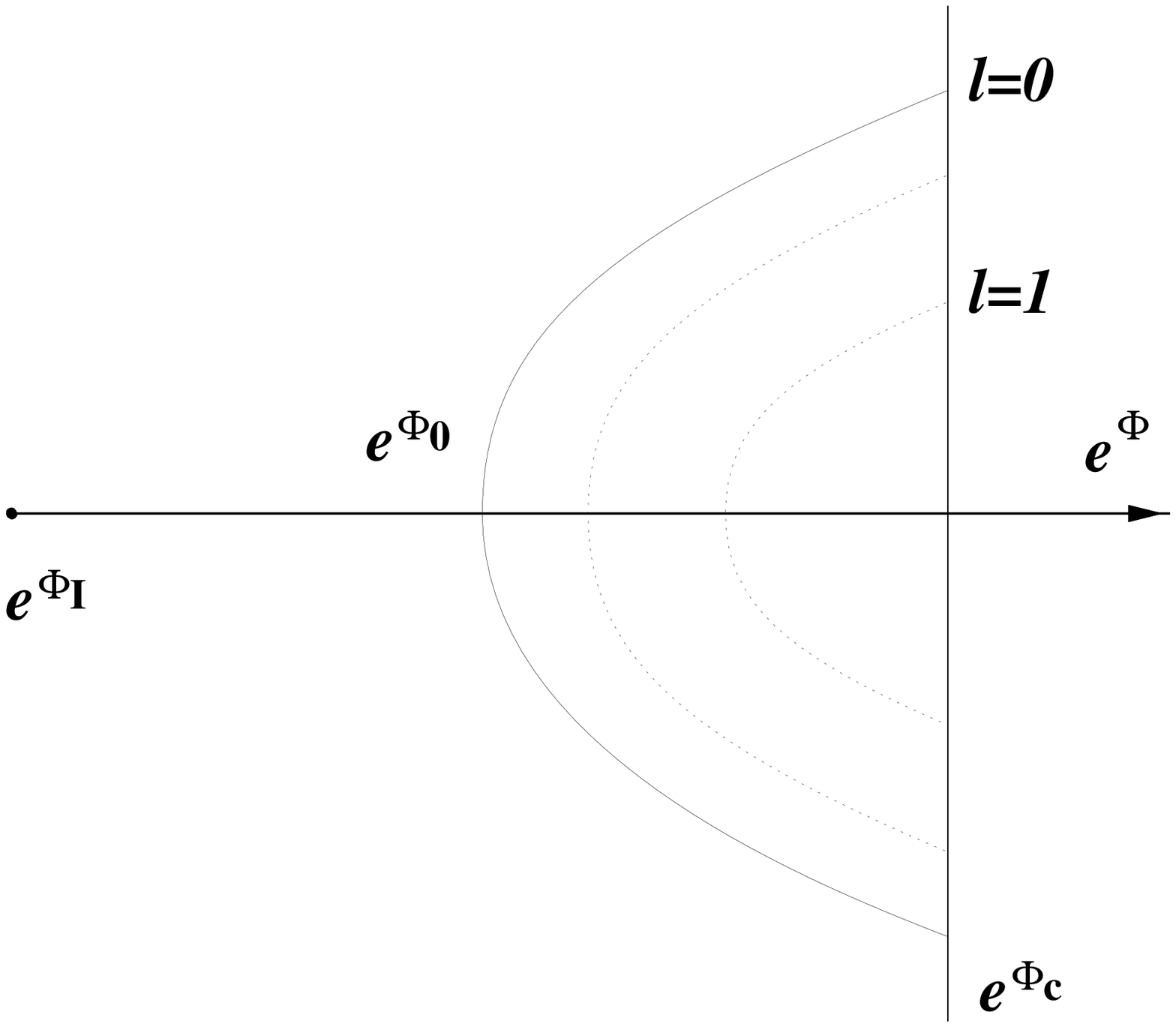, width=9cm}
\end{center}
\vspace{0.5cm}
\baselineskip=13pt
\centerline{\small{Figure 5: Schematic shape of the critical minimal}}
\centerline{\small{surfaces used in calculating the Wilson-loops.}} 

\vspace{0.5cm}

\baselineskip=20pt plus 1pt minus 1pt

Therefore, there are effects due to the flat extra dimensions on the separation
$L$ and also on the string tension. The length of the Wilson loop $L$ 
(measured on the Minkowski spacetime) decreases as the parameter $l$ increases. 
On the other hand, the separation of the {\it quark-antiquark} pair
increases in the hyper-plane ${\cal{H}}$. The string tension also increases by a factor
$\frac{1}{\sqrt{1-l^2}}=({\cos \alpha})^{-1}$, as it is expected. \\

Concerning the singularity and its consequences on the dual field theory,
in an analogous situation as the one described in \cite{ALL}, the existence of a maximum for
$L$ in terms of $e^{\Phi_0}/e^{\Phi_c}$, as we mentioned before, implies the
existence of two regions. This phenomenon resembles the case studied in reference
\cite{REY} related to a Schwarzschild-anti de Sitter solution. In such a situation
a horizon appears and  covers the singularity. 
Taking into account \cite{REY}, in our case it has been suggested
\cite{ALL} that the point $e^{\Phi_0^M}/e^{\Phi_c}$, where $L$ has a maximum, 
could be interpreted as an effective horizon. This is equivalent to say that there is 
a cut-off preventing to reach the singularity.
If one trust this effective horizon, then only in region II it is possible to draw Wilson loops.
This means that from the side of the gauge theory
we should have only confinement, and not over-confinement. Nevertheless, even when we do not
consider that Schwarzschild-anti de Sitter case of reference \cite{REY}, 
the singularity which is present in our construction
is not contained in the usual classification scheme \cite{SINGULARITY} since we have not
a scalar potential in the equations of motion, as that criterion requires, in order to decide
whether the singularity is good or bad. On the other hand, one can try to analyze 
the situation presented here under the light of the criterion of reference \cite{CHARLES}.
In that reference the related issue is that the time-like component  of the metric $g_{00}$
has to be either a non-increasing function or it should be bounded from above as one
approaches the singularity. These are the strong
and weak versions of the same criterion, respectively. In fact, in our case $g_{00}$ is zero at the singularity,
in this sense we could think that the solution is acceptable although we can not 
trust the dual gauge theory if we analyze a region near the singularity because the spacetime 
becomes strongly curved as we can see from the scalar curvature. Taking into account these
arguments we can consider that the way to deal with our singularity should be the case
suggested in \cite{ALL} by understanding that a cut-off has to be introduced around the singularity.
On the other hand, in relation to the criterion of \cite{CHARLES} we should also be careful
in the case when $\rho \rightarrow \infty$, because from Eq.(4) $g_{00} \rightarrow \infty$
as  $\Phi \rightarrow +\infty$. This, of course, corresponds to the strongly coupled limit
of the gauge theory. Nevertheless, this is not a problem because, in principle,  we can 
exclude the limit $\Phi \rightarrow +\infty$ in the present analysis so that the results 
presented here remain valid. \\

We would like to thank Ju\'an Pedro Garrahan, Ian Kogan, Bayram Tekin, Michael Teper for useful discussions.
We thank Carlos N\'u{\~n}ez for a critical reading of the manuscript and his useful comments, as
well as Rafael Hern\'andez for correspondence. 
This work has been partially supported by the CONICET of Argentina, the Fundaci\'on Antorchas 
of Argentina and The British Council. 


\vfill

\end{document}